\documentclass[conference]{IEEEtran}
\IEEEoverridecommandlockouts
\usepackage[hyphens]{url}

\usepackage{subcaption}
\usepackage{cite}
\usepackage{amsmath,amssymb,amsfonts}
\usepackage{algorithmic}
\usepackage{graphicx}
\usepackage{textcomp}
\usepackage{xcolor}
\def\BibTeX{{\rm B\kern-.05em{\sc i\kern-.025em b}\kern-.08em
    T\kern-.1667em\lower.7ex\hbox{E}\kern-.125emX}}
\begin{document}

\title{Blockchain-Based Secure Vehicle Auction System with Smart Contracts\\}

\author{\IEEEauthorblockN{Ka Wai Wu}
\textit{Virginia Tech}\\
kawai@vt.edu \\
USA
}

\maketitle

\begin{abstract}
The problem of a single point of failure in centralized systems poses a great challenge to the stability of such systems. Meanwhile, the tamperability of data within centralized systems makes users reluctant to trust and use centralized applications in many scenarios, including the financial and business sectors. 

Blockchain, as a new decentralized technology, addresses these issues effectively. As a typical decentralized system, blockchain can be utilized to build a data-sharing model. Users in a blockchain do not need to trust other users; instead, they trust that the majority of miner nodes are honest. Smart contracts enable developers to write distributed programs based on blockchain systems, ensuring that all code is immutable and secure.

In this paper, we analyze the security of blockchain technology to illustrate its advantages and justify its use. Furthermore, we design a new system for storing and trading vehicle information based on the Ethereum blockchain and smart contract technology. Specifically, our system allows users to upload vehicle information and auction vehicles to transfer ownership. Our application provides great convenience to buyers and owners, while the use of smart contracts enhances the security and privacy of the system.

\end{abstract}

\begin{IEEEkeywords}
Blockchain, Smart Contract, Ethereum, Bid, Solidity
\end{IEEEkeywords}

\section{Introduction}
\subsection{Problem Motivation}
In recent years, as the interaction between people, things and the
environment becomes more frequent, more and more information is created and broadcasted. In order to save energy and improve efficiency, different objects share data through the network. Data needs to be transferred and stored over the network, and programs are needed to control it. For example, in a used vehicle trading application, we need to store the vehicle id, owner, year, and price estimate, and we need to provide an auction venue for other users to purchase the vehicle to obtain ownership. All such information should be publicly available and transparent, and users should be successfully authenticated before information can be added, updated, and deleted. 

Our traditional models use centralized servers to conduct the
service. For example, some dealer websites can help users enrich
the information on used cars for auction. Although this mode can
save some cost for storage and processing, it has some security limitations that need to be considered. Firstly, the centralized servers have all the permissions to store and transfer data. Managers cannutilize the data for illegal purposes. According to the description of the PRISM program \cite{5}. people’s data stored by some Internet and telecommunication companies has been exploited, which means that our data will be modified in the systems by those third parties. Secondly, data stored in centralized servers cannot, in principle, be guaranteed to be unchangeable. Therefore centralized servers
may collect more information about users to counteract data modification. Whether the data is altered for external reasons or by the system itself, users do not want to trust these servers in certain
scenarios \cite{li,wu}. For example, the trading platform for used cars may
crash due to some problems causing users to not be able to use it
for a while. In addition, private tampering of used car information
is also possible, which can generate a lot of losses.

Blockchain can be a good solution for us to solve those problems.
Since bitcoin’s inception, blockchain technology has piqued great
interest from the industry and application providers. Blockchain is
a decentralized ledger that can process and store information like
transactions. Blockchain can keep the data stored there and not
be modified because of the proof of work and the honest miners’
computing power. As for anonymity, blockchain technology uses
public-key cryptography to identify accounts’ addresses. Nobody
knows who owns one account. Therefore, one user can utilize different accounts to transfer data and keep the owner secret. Network structures and connectivity play a crucial role in decentralized systems, as highlighted in \cite{sung2025community}, which can help optimize transaction flows and community-based trust mechanisms in blockchain applications. As for
considering the blockchain technology in the data-sharing systems,
the resource owner can feel free to upload and share their resource
without concerning with many external factors. Furthermore, each
node in the blockchain is equal and shares the same function which
means that everyone could be the resource sharer and customer. It is convenient for the blockchain user to share and buy the resource.
Besides, the emergence of smart contracts also provides a decentralized programming model for handling data in the blockchain, allowing users to handle data more easily. Smart contracts make it impossible to tamper with the flow processing of data and control the conditions.

\section{Key Ideas}
Considering the characteristics of the smart contract, the key idea
of this project is to put digital properties into the blockchain, and
all access to the properties is conducted by the smart contract. More
specifically, taking the second-hand marketplace as an example,
users can post the goods they want to sell on the blockchain, including information about the goods they want to sell and the expected
selling price. Users can use the program to control the auction of
goods and the flow of money. In this way, all transactions will be
honest and cannot be tampered with. Once the goods have been
traded, the information about the goods should be updated in time
for the next transaction. Blockchain information is open and transparent so that everyone can see the information about the goods. Additionally, natural language processing techniques such as sentiment analysis and topic modeling \cite{sung2025nlp, sung2025housing} can enhance user interactions in blockchain-based marketplaces by extracting meaningful insights from transaction data and customer feedback. The implementation of this technology in the physical world needs
to be further extended because the delivery of goods in the physical world requires a realistic face-to-face process, and the problem
solved by blockchain technology can only be the completion or
failure of a transaction in a procedural sense and does not ensure
the delivery of physical goods. However, this technology has a wide
range of applications in the Internet of Things and other network
technologies. This is because the goods delivered are also virtual
products based on the web.

\subsection{Proposed contributions}
Firstly, we design and implement a new application of used car
bidding, using the smart contract. All the bidding process and car
info management are maintained by the smart contract, ensuring
its privacy and safety. Compared with existing used car trading
applications, our application takes advantage of the security and
privacy of the smart contracts, ensuring the safety and privacy of
the transactions between users. Besides this, all the records of a car
are updated by the agent, ensuring the trust between users.

\section{ANALYSE OF BLOCKCHAIN}
Blockchain, as a decentralized ledger, is not only a normal database
when we consider how to use it. We need to point out its unique
advantages compared to traditional centralized servers.

\subsection{Integrity}
As for the definition of integrity \cite{1}, it is the degree to which a
system prevents unauthorized access to, or modification of, computer programs or data. If a system has high integrity, the system
usually has strong mechanisms to protect data from modification.
In centralized systems, the third-party servers are needed to be
trusted. Then those servers can deploy other security mechanisms
to protect data and activities when they are transferred and processed by servers. However, if users do not want to trust centralized
servers, everything the server does is in vain. It means that the servers themselves are likely to modify data for invalid purposes,
which will destroy the integrity of the systems.

As for the blockchain system, it is built by thousands of nodes
in the network. In this system, the key to protecting the integrity
of the system is that nodes with the most capabilities are honest
ones. They will create valid blocks and verify others’ new blocks.
Most blockchain consensus is based on proof of work. Miners need
to solve a hard math problem, which means getting the result
of the nonce value according to data of the previous block and
transactions’ information, to create a new block. The nonce is hard
to calculate but easy to verify. As for the proof of work, someone
who calculates the nonce value firstly can add the block in the
blockchain, where the block is verified by others, which means
that the data in the block is also valid. Therefore, as for the way to
destroy the integrity of the blockchain system, valid data need to
be covered by invalid data and the system will not be trusted by
participants. The rule about how to identify the valid block is based
on the longest valid chain. Attackers with high computing power
may fork the blockchain and if their private branch is longer than
the normal chain, their attack will be successful and they can add
some malicious data to their chain. The best method is choosing
a stable blockchain, where each miner will not have the too high
computing power \cite{4}.

\subsection{Confidentiality}

In a blockchain system, the confidentiality of data is weak \cite{lish,sok}. The
reason is that all data stored in the blockchain system is public,
including transaction information and smart contract code and
smart contract status values. As we mentioned earlier, all the data
in the blockchain is backed up in all the full nodes so that it can be
counteracted against a single point of failure. In addition to this, all
new data on the blockchain is based on transactions, or on changes
in the state of the contract account generated by contract calls. The
transactions need to be extracted from the transaction pool by the
miners, who obviously need to know the content of the transactions
in order to update the data in the local tree data structure and verify
whether the transactions are legitimate. In addition, when a newly
generated block is propagated to other nodes, the honest nodes need
to verify the legitimacy of the block transaction, so the transaction
data must be made public for other nodes to query. Therefore, the
openness of transactions is a prerequisite to ensure the blockchain
as a fully decentralized system.

Ekiden\cite{3} might be a method for improving blockchain confidentiality. It is used to connect the blockchain and TEE. Ekiden
provides a safe platform for smart contracts. Ekiden may be thought
of as a state machine. We can acquire the new state and output by
using the previous state and the user’s input. Privacy and secrecy
can be guaranteed if these items are used in the Ekiden. In the Ekiden, there are three types of creatures. The primary consideration
is clients. They are, in fact, nodes that leverage contracts. These networks can both develop and execute smart contracts. The clientele
are light, and the computation is not carried out. The second computes nodes capable of processing client requests by running the
contract in TEE and generating attestations verifying that the state
modifications are valid. The last nodes are those that have reached
an agreement. They feature a distributed append-only ledger that solely runs the blockchain’s consensus process. The workflow is
the next step. The first step is to draft a contract. The contract code
must be sent to the computing node by the client. The compute
node then loads the contract into a TEE and initializes it. Then,
in the initial state, a new contract id and the necessary keys from
key management are produced and encrypted. By contacting the
attestation service, the compute node acquires verification of the
validity of the state. The following step is to request execution. The
node must do some calculations and interface with the blockchain
system. In terms of security performance, the system can impose
some required security qualities. The first need is proper execution.
The change in the state of the contract demonstrates the accuracy
of the execution. The second point to consider is consistency. The
blockchain must achieve a consensus state sequence. Furthermore,
the system may get the secret without taking into account the assault or failures. If there is indeed a breach of confidentiality, there
will be isolation.

\subsection{Adaptability}
As for the blockchain system, there are two adaptability issues. The
first one is that with the increasing number of transactions, the
blockchain enlarges its size, which means that it becomes more
expensive to store the data of the blockchain, especially when some
light clients do not have so much space.

The second one is the data media throughput, and transactions
in the blockchain. In the bitcoin system, the size of the block is only
1m, and creating a new block needs about 10 minutes, which is
too slow. But in Ethereum, the block-creation time can be adjusted
according to the request. Normally, the time is about 15 seconds.
There much more transactions that can be stored. Therefore, based
on adaptability, we choose Ethereum as our application platform.
Because our application clients are not able to conduct mining
activities or store all block data as full nodes, they only need to
initialize transactions including sensitive data in the Ethereum net-
work. Other miners will select those transactions in the blockchain.
Users need to pay the gas fee for blockchain’s service.

\section{APPROACH OVERVIEW}
To develop our application, we surveyed the smart contract about
what it is, how it works, and how we could develop an application
using this technology. After all the investigation, we decided to
develop our application on the Ethereum platform using solidity,
and it is what most smart contract applications do. Ethereum is an
open-source platform that builds decentralized applications, and
solidity is an object-oriented, high-level language for implementing
smart contracts. Besides this, we will use the remix platform as the
IDE for our application. Ropsten network as our test environment.
We use the injected Web3 to interact with the smart contract to test
our application.

\section{DESIGN}
\subsection{System Architecture}
\begin{figure}[htbp]
    \centerline{
    \includegraphics[width=0.5\textwidth]{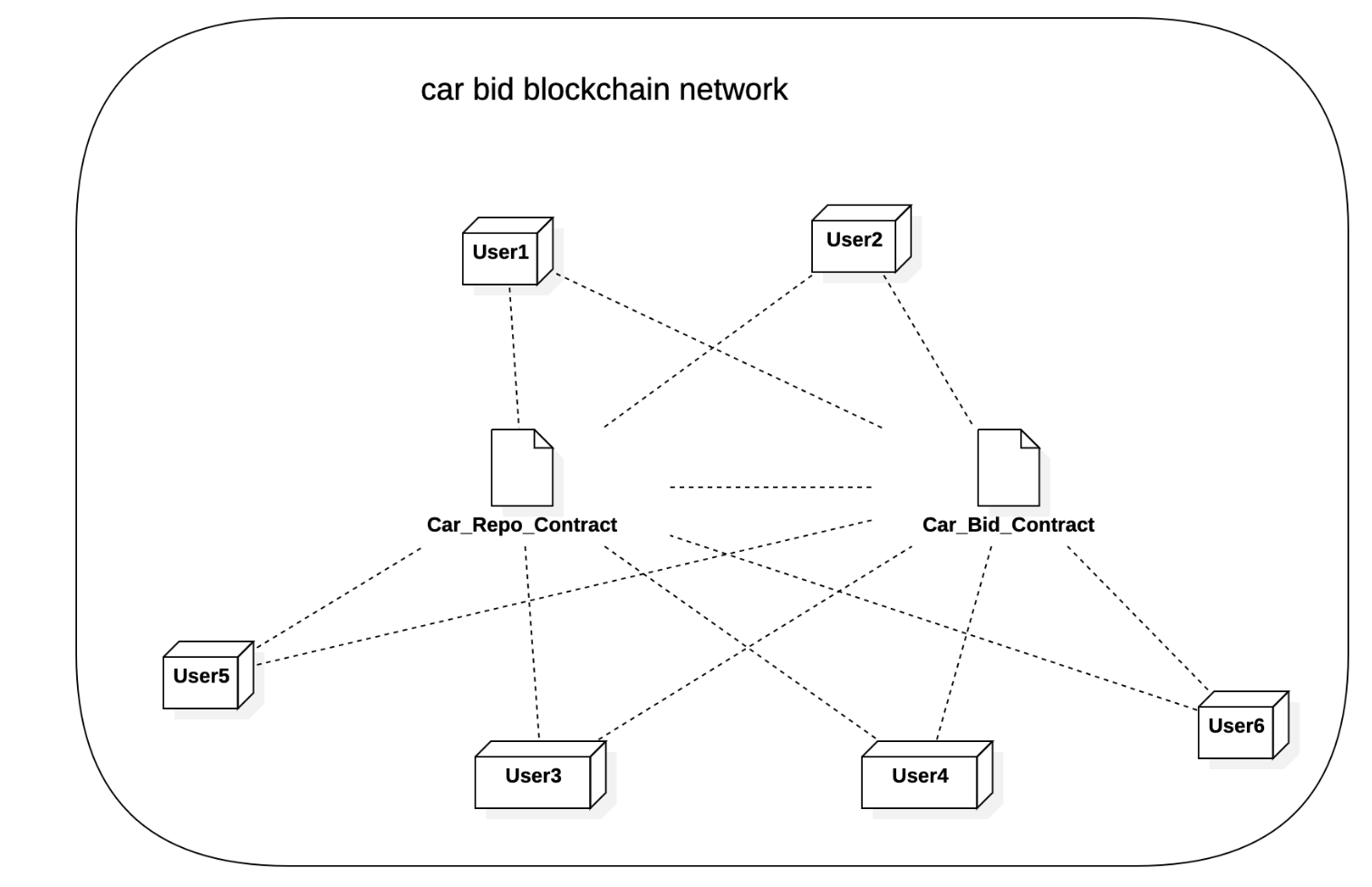}}% Replace with the path to your image file
    \caption{System Design}
    \label{fig:fig1}
\end{figure}

Figure~\ref{fig:fig1} shows our overall system architecture. It’s a peer-to-peer blockchain network.

Two main contracts are running in our system, the car repo contract, and the car bid contract. Every user running in our network
will obey the rules of the two contracts.

The car repo contract is used for managing the cars. The car
agent can manage the cars being bid using this contract. It contains
a list of cars that are available for sale.

The car bid contract is mainly used for buyers to bid. When a
buyer selects a car and offers the price, this contract is responsible
for determining whether the bid is valid or not, whether the bid
time is still available, etc.

\subsection{Use Case Design}
The users of our system can act as the car agent, the buyer, and the
car owner.

\begin{figure}[htbp]
    \centerline{
    \includegraphics[width=0.5\textwidth]{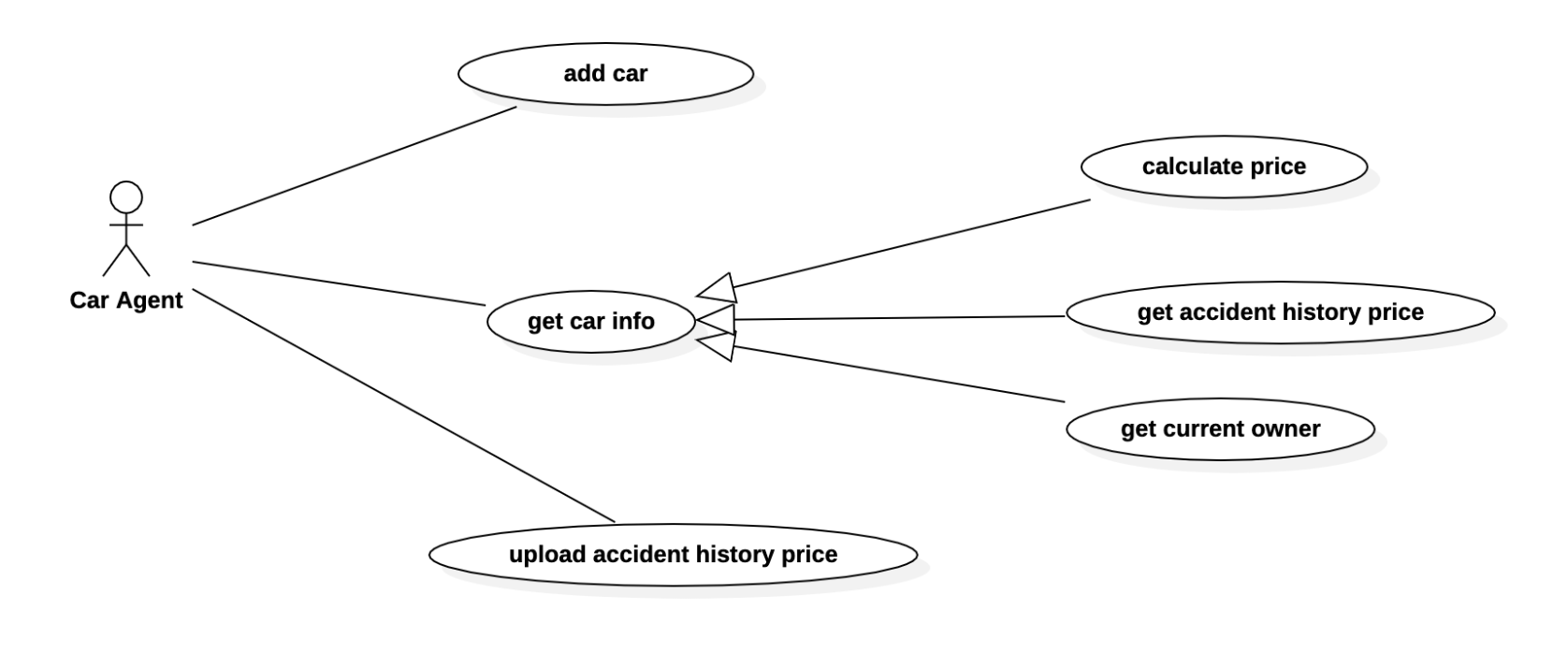}}% Replace with the path to your image file
    \caption{Use Case For Car Agents}
    \label{fig:fig2}
\end{figure}

The car agent (see Figure~\ref{fig:fig2}) is responsible for managing the cars
being bid. They can add the cars for bidding, get the information
of the cars including its estimated cost, their accident history fee,
and get the current owner of the cars. Besides this, the car agent
should be responsible for uploading the accident history fee of a
car in order to calculate the estimated price for the cars.

\begin{figure}[htbp]
    \centerline{
    \includegraphics[width=0.5\textwidth]{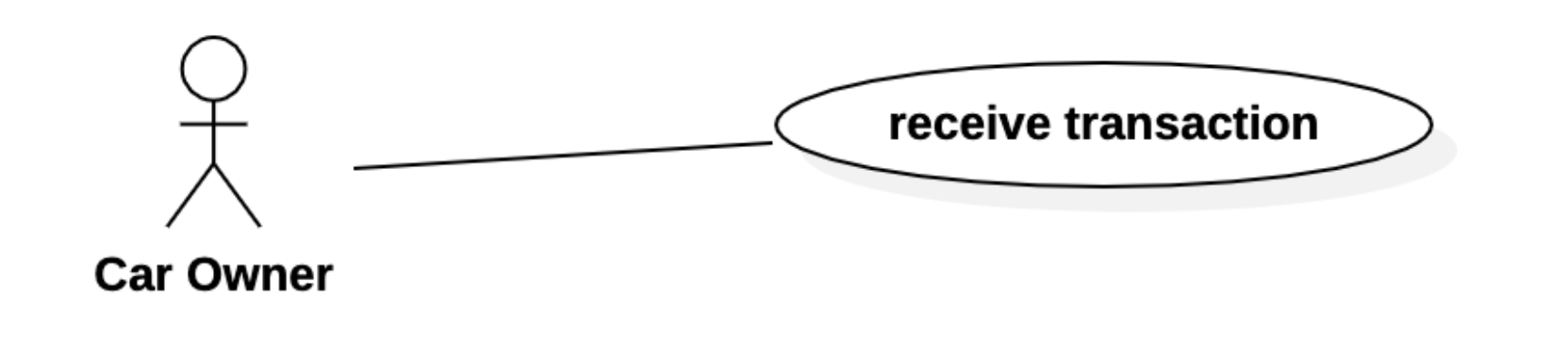}}% Replace with the path to your image file
    \caption{Use Case For Car Owners}
    \label{fig:fig3}
\end{figure}

The car owners (see Figure~\ref{fig:fig3}) only need to receive the money as
long as their cars are being bought by the buyers.

\begin{figure}[htbp]
    \centerline{
    \includegraphics[width=0.5\textwidth]{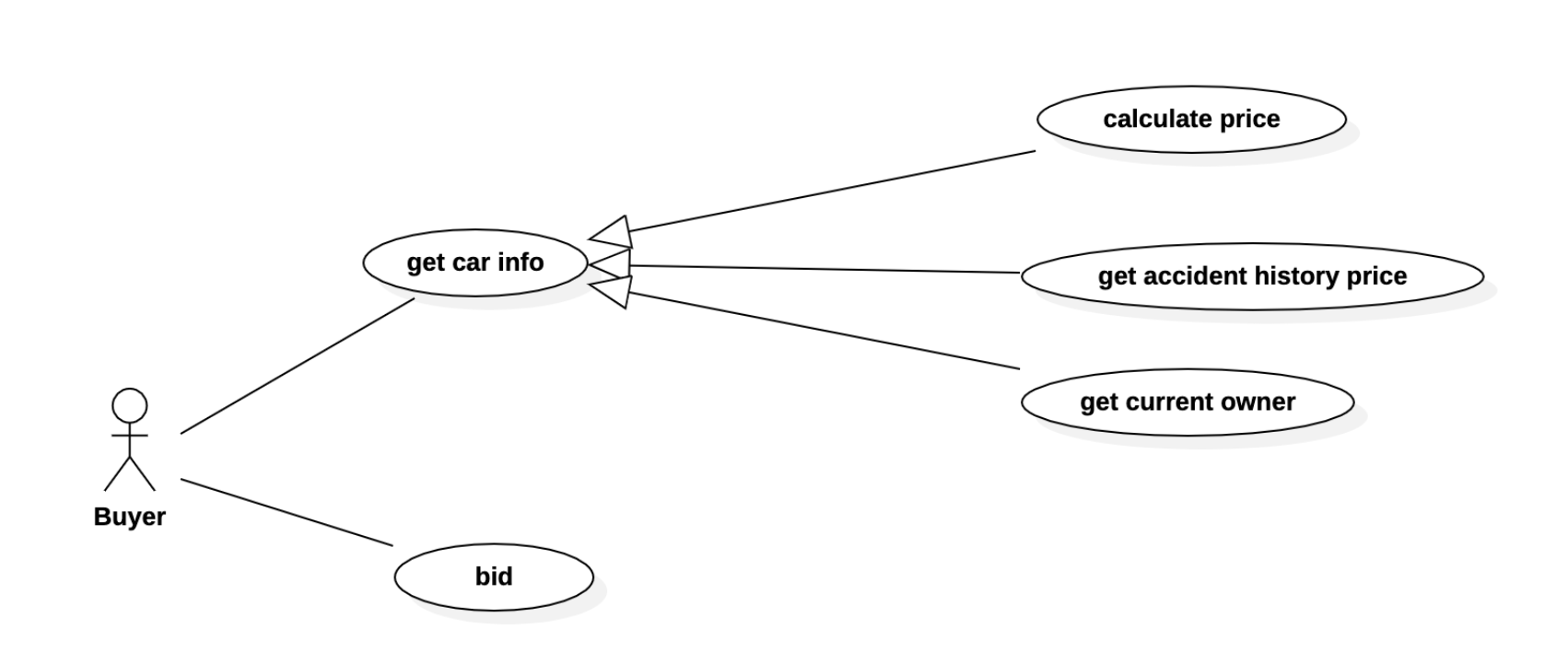}}% Replace with the path to your image file
    \caption{Use Case For Car Buyers}
    \label{fig:fig4}
\end{figure}

The buyers (see Figure~\ref{fig:fig4}) can choose the car they are willing to
buy to get the estimated price, the accident record of the car and its
current owner. As long as the buyer makes his decision, he can offer
their price for the car and wait for the system to decide whether he
is the highest bidder to get the car or not. When he wins the bid,
the system will send its money to the car owner.

\subsection{Class Diagram}
\begin{figure}[htbp]
    \centerline{
    \includegraphics[width=0.5\textwidth]{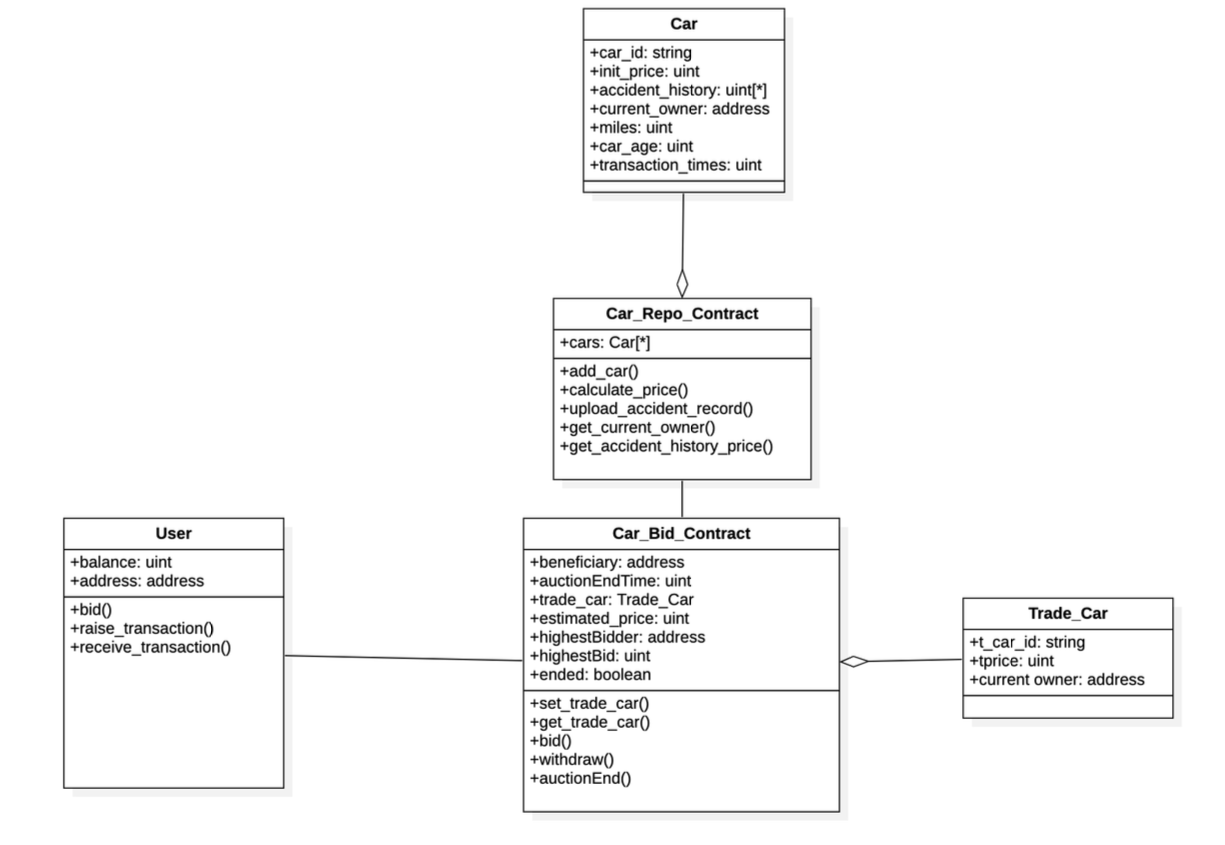}}% Replace with the path to your image file
    \caption{Class Diagram}
    \label{fig:fig5}
\end{figure}

According to Figure~\ref{fig:fig5}, Our system has 5 classes, which are User,
Car, Car\_Repo\_Contract, Trade\_Car, and Car\_Bid\_Contract respectively. The Car class is an inner class inside the Car\_Repo\_Contract,
which is the trading entity during the transaction. A car has its
specific id, current owner, initial price, and other properties like
miles, ages, accident history, etc. The Car\_Repo\_Contract class is
the smart contract that maintains a map struct called cars with the key of car\_id, the value of the Car entity. Once there is a car
under the hammer, the car agent will input the car\_id to retrieve the car instance from the cars map. Besides, inside the car\_repo smart
contract, there are 5 functions to manipulate the Car transactions.
We can add cars to the map of the car, upload specific car accident
history costs (for example, if there is an accident happened to the
car, the car agent could estimate the discount of the car’s initial
price and add to the accident history costs), get the current owner
of a specific car, and most importantly, calculate car price for a later
bidding system. The calculate\_price function will calculate the current car sale price based on the car’s ages, miles, accident\_history,
and trade times. Then comes the bidding part. There is a User class
that participates in the bidding process. The User has its balance,
address, and functions like bid, raise\_transaction, and receive transactions (because a user could be the current owner of the car), and
withdraw their bids since the bidding price may exceed their budget. The next class is Trade\_Car, which is an inner class inside the
Car\_Bid\_Contract, it is a struct that contains the car information
for which will be traded. The Trade\_Car has a property of tprice,
which is the estimated price calculated by the Car\_Repo Contract
calculate\_price function, and sent back to the Car\_Bid\_contract.
During the auction procedure, the bidder bid with a price lower
than tprice will be directly refused. The Car\_Bid\_Contract class has
properties like beneficiary, auctionEndTime, trade\_car, etc. And
functions like bid, set\_trade\_car, withdraw, etc. there is an auction
end time set to limit the auction time.

\subsection{Algorithm Design}

\begin{figure}[htbp]
    \centerline{
    \includegraphics[width=0.5\textwidth]{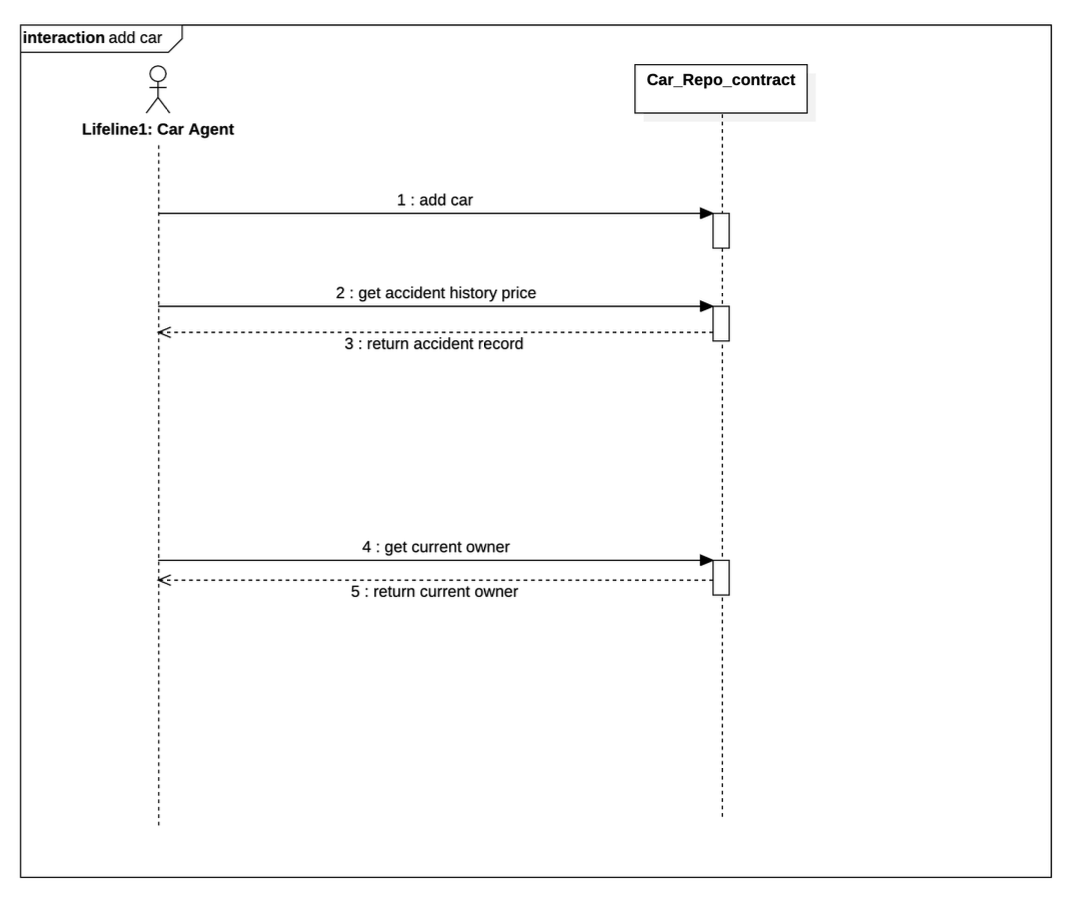}}% Replace with the path to your image file
    \caption{Add Cars}
    \label{fig:fig6}
\end{figure}

During this process (Figure~\ref{fig:fig6}), when there is a new car for sale,
the car agent can add the car into the cars map by setting the car’s
basic information like age, initial price, miles, etc. Then the car agent
can acquire the car’s information from the Car\_Repo\_contract like
accident history, current owner.

\begin{figure}[htbp]
    \centerline{
    \includegraphics[width=0.5\textwidth]{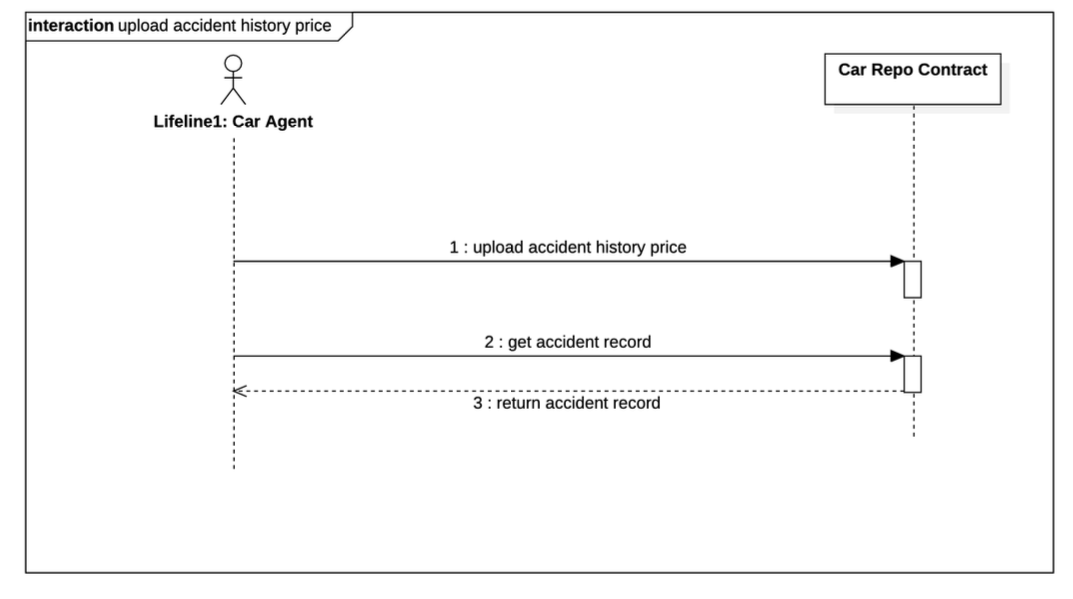}}% Replace with the path to your image file
    \caption{Upload Accident History Price}
    \label{fig:fig7}
\end{figure}

According to Figure~\ref{fig:fig7}, the Car agent can upload the car accident
history price to the Car instant, and the agent can get the accident
record of the car.

\begin{figure}[htbp]
    \centerline{
    \includegraphics[width=0.5\textwidth]{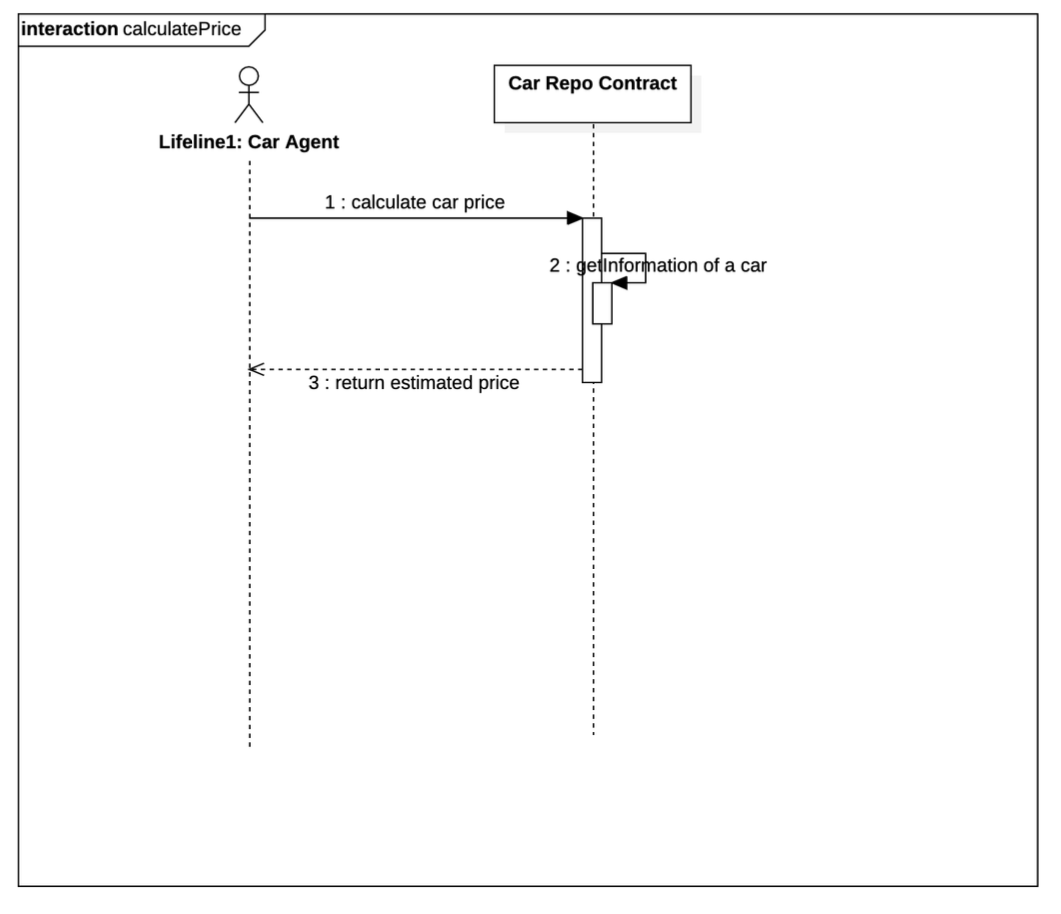}}% Replace with the path to your image file
    \caption{Calculate Price}
    \label{fig:fig8}
\end{figure}

The Car agent can calculate the car price (Figure~\ref{fig:fig8}) and get
an estimated price for a later auction. The calculate function will
retrieve the car instance and read its information. Based on the
information like age, miles, accident history, and transaction times,
it will make a calculation and return back an estimated price.

\begin{figure}[htbp]
    \centerline{
    \includegraphics[width=0.5\textwidth]{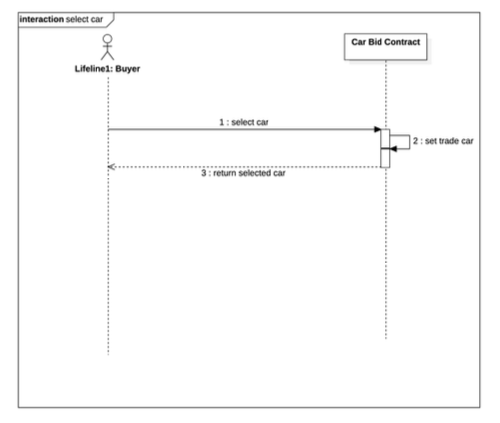}}% Replace with the path to your image file
    \caption{Select a Car}
    \label{fig:fig9}
\end{figure}

During the Auction procedure, a car will be selected (Figure~\ref{fig:fig9})
for bidding. It will return the specific trade car id.

\begin{figure}[htbp]
    \centerline{
    \includegraphics[width=0.5\textwidth]{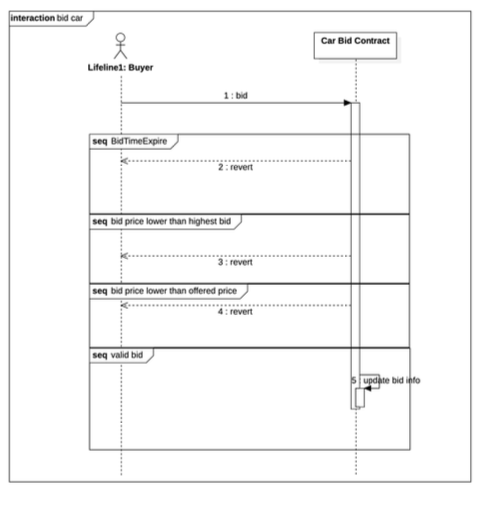}}% Replace with the path to your image file
    \caption{Bid a Car}
    \label{fig:fig10}
\end{figure}

The buyers will bid on the car (Figure~\ref{fig:fig10}). Each of the buyers will raise a transaction to the Car Bid contract. There are 3 conditions
for bidding on a car, first, if the bid time has expired, the bidding
transaction will be directly reverted. Secondly, if the bid price is
lower than the highest bid price, it will also be reverted. Then
if the bid price is lower than the initially offered price, which is
calculated by the Car\_Repo\_contract, the transaction will also be
directly reverted. Once the bid transaction is valid, the least bid
transaction will be recorded and updated into the Car Bid contract.

\begin{figure}[htbp]
    \centerline{
    \includegraphics[width=0.5\textwidth]{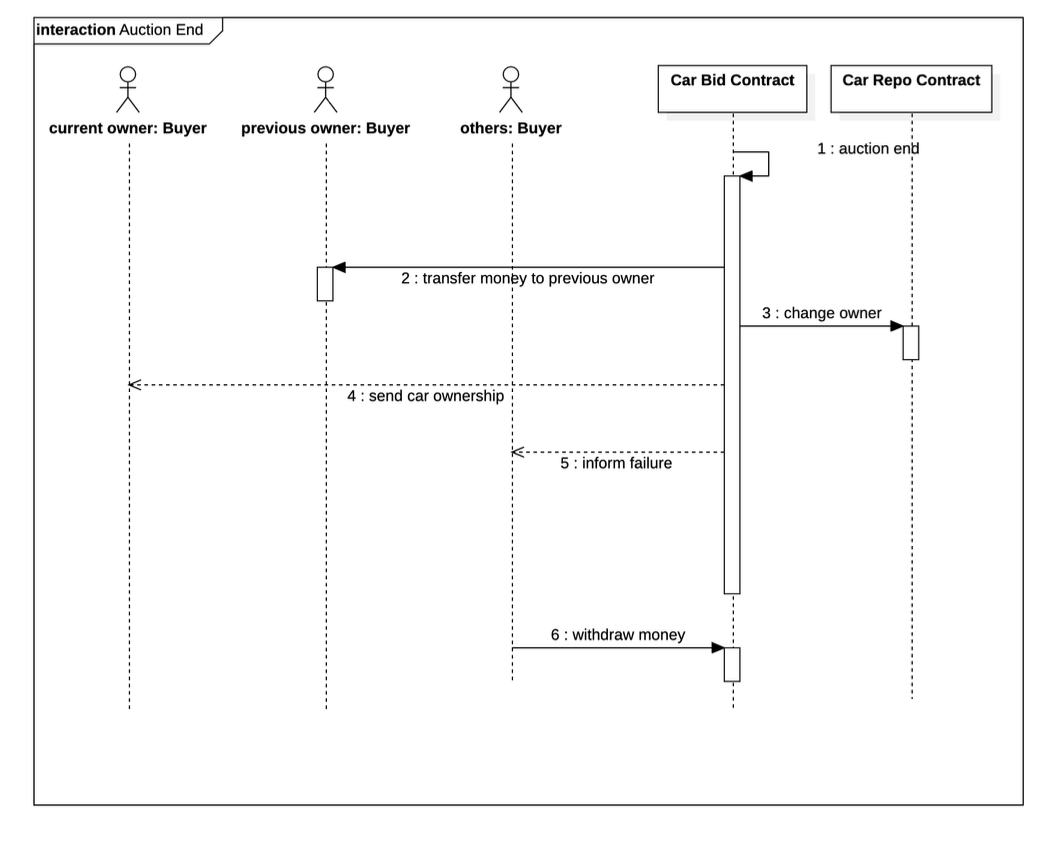}}% Replace with the path to your image file
    \caption{Auction End}
    \label{fig:fig11}
\end{figure}

The system needs to judge whether the auction ends or not (Figure~\ref{fig:fig11}). If the timestamp of the block is still earlier than the scheduled auction end time, then the autionNotYetEnded event
is revoked. If the auction is already closed, revoke the event of
AuctionEndAlreadyEnd. If it is time to end the auction, the system
will transfer money from the previous owner to the buyer, and
change the owner of this car, and tell the new owner he wins the
auction. The system will also tell the other bidders that they did not
win the auction, the other bidders then can withdraw their money
from the system.

\subsection{Implementation}
During the implementation, we coded with Solidity with version
0.8.4 to build up the system, under the Remix online IDE platform.
In the Ethereum test network, we chose the Ropsten test network
to mine and place transactions. The digital wallet we used is MetaMask.

The Car\_Repo\_Contract uses mapping as the structure for managing cars available for sale. Several standards are applied to calculate the estimated price, ranging from the car's age to its mileage and accident history. In the end, a final price for the car is determined based on these factors.

In the Car\_Bid\_Contract, an inner class called Trade\_Car is used as the entity to represent the bidding goods. The instance of the trade car can be accessed through the Trade\_Car entity. During the bidding process, if a bidder with a proposed bidding price is not qualified, the transaction is directly reverted. Otherwise, the bidding transaction is recorded, and the highest bidding price along with the bidder's information is updated.

\section{Evaluation}
\begin{figure}[htbp]
    \centerline{
    \includegraphics[width=0.5\textwidth]{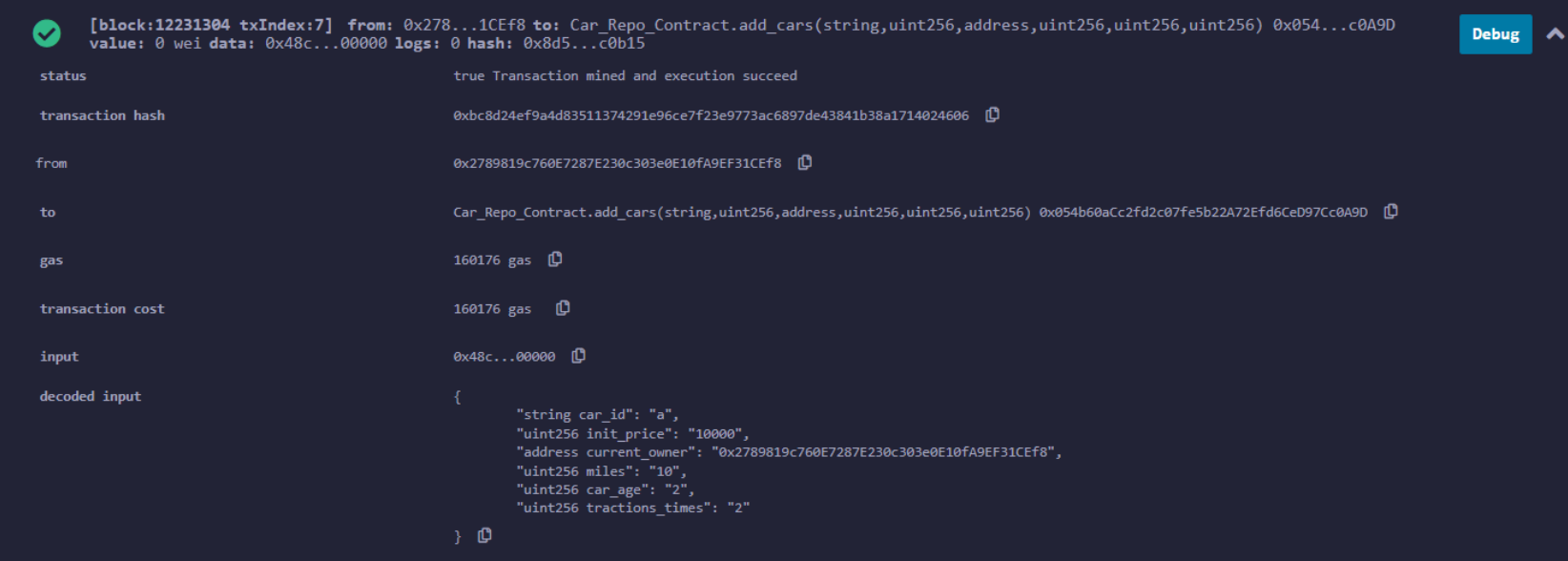}}% Replace with the path to your image file
    \caption{Make a Transaction}
    \label{fig:fig16}
\end{figure}

\begin{figure}[htbp]
    \centerline{
    \includegraphics[width=0.5\textwidth]{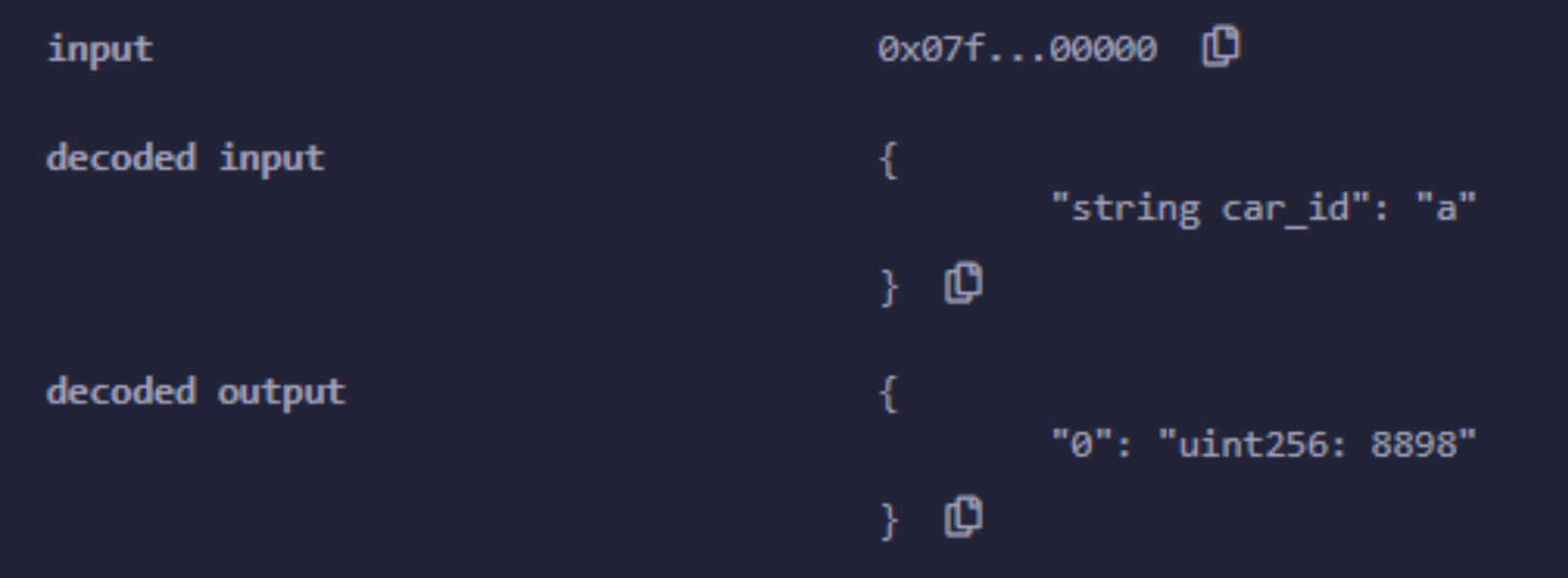}}% Replace with the path to your image file
    \caption{Calculation}
    \label{fig:fig17}
\end{figure}

We successfully made a transaction to add a new car to the cars list
with the default value (Figure~\ref{fig:fig16}).
For calculating the estimated price, we select the car id of which
the car we want to calculate, then we got the price 8898 after
calculation (Figure~\ref{fig:fig17}).

\begin{figure}[htbp]
    \centering
    \begin{subfigure}[t]{0.5\textwidth}
        \centering
        \includegraphics[width=\textwidth]{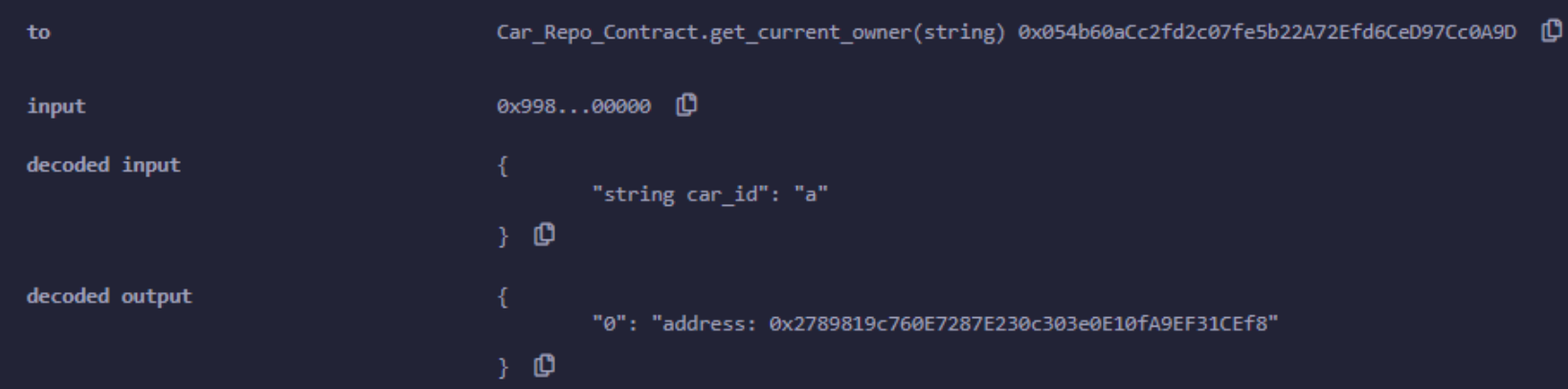}
        \caption{Initial state}
    \end{subfigure}%
    \\[0.5em]
    \begin{subfigure}[t]{0.5\textwidth}
        \centering
        \includegraphics[width=\textwidth]{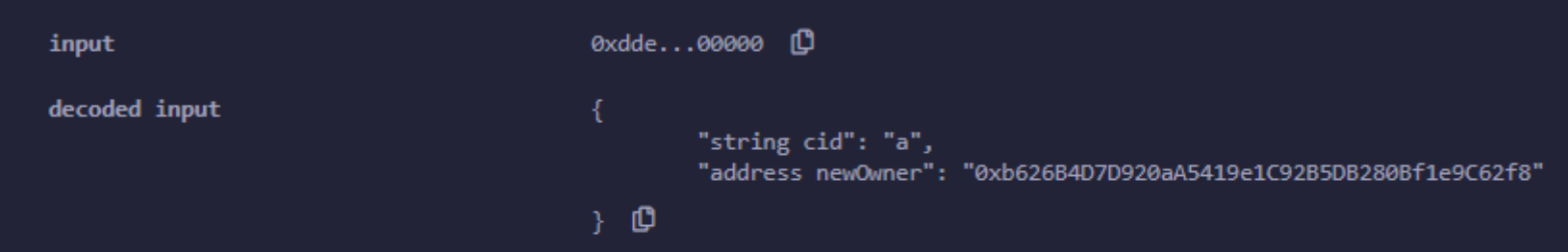}
        \caption{Updated state showing the new owner address}
    \end{subfigure}
    \caption{Change Owner Function}
    \label{fig:change_owner}
\end{figure}

We tested the change owner function and got a new owner with its address (Figure~\ref{fig:change_owner}).

\begin{figure}[htbp]
    \centering
    \begin{subfigure}[t]{0.5\textwidth}
        \centering
        \includegraphics[width=\textwidth]{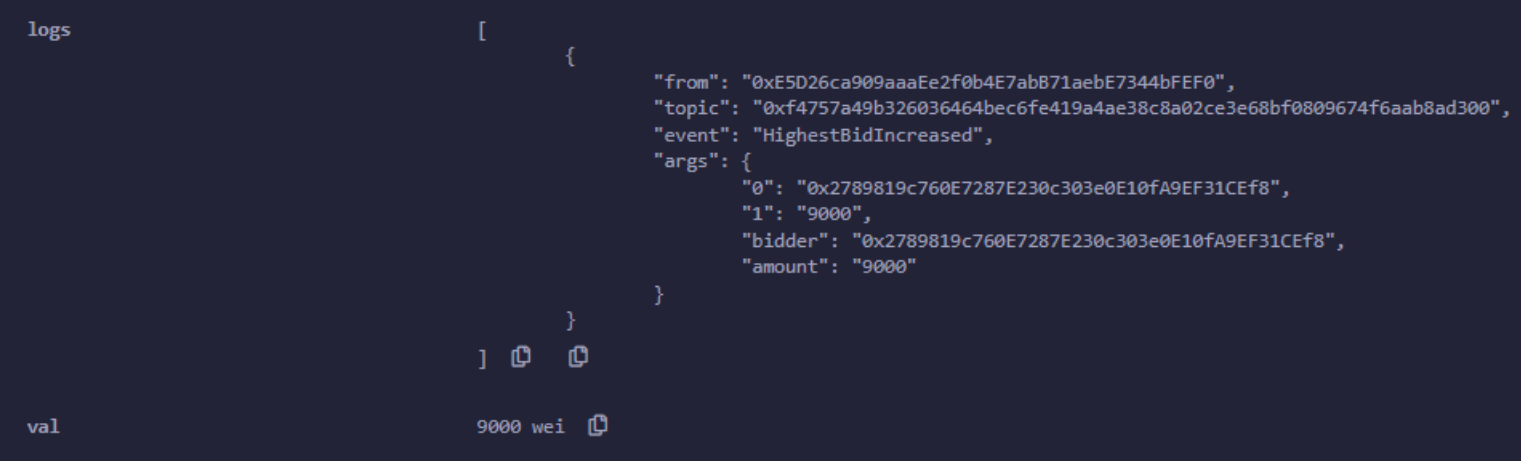}
        \caption{Initial bid record}
    \end{subfigure}%
    \\[0.5em]
    \begin{subfigure}[t]{0.5\textwidth}
        \centering
        \includegraphics[width=\textwidth]{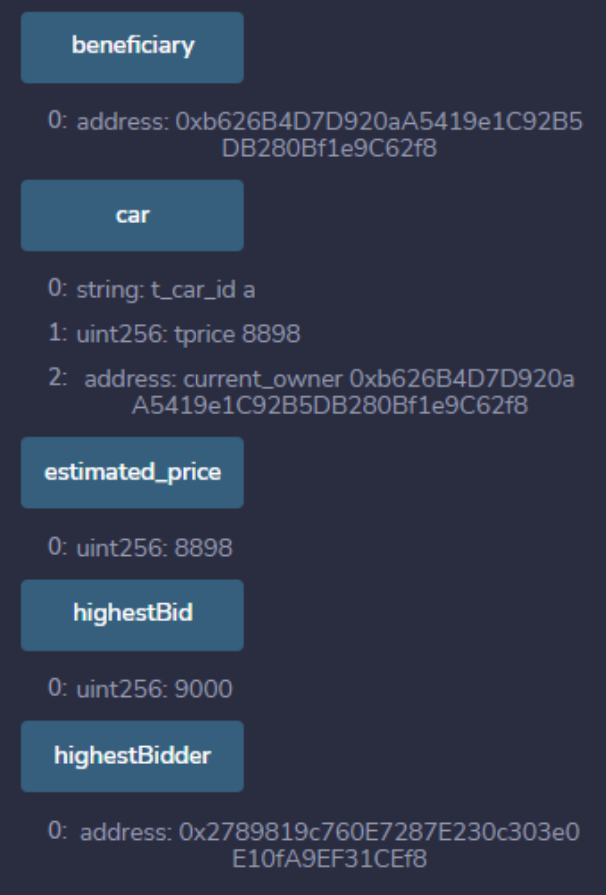}
        \caption{Updated record showing additional bids}
    \end{subfigure}
    \caption{Bidding Process}
    \label{fig:bid_process}
\end{figure}

During the bidding process, we made several bids, and the auction record was updated correspondingly (Figure~\ref{fig:bid_process}).

\textit{Limitations.} Due to time as well as knowledge limitations, there
are some problems with the system we designed. First, we ignored
some design when making conditional judgments. Different understanding of some judgments in smart contracts can cause different results\cite{6}, and our code may have ignored these issues. The reason is that we are not familiar enough with the underlying architecture of the smart contract code. Secondly, our code is missing the frontend design and is not a complete dApp. we just write the backend code of this used car auction system using remix platform. We will improve our code after the course and upload it to Github. Finally, our algorithm is high in memory and other resource consumption, and we did not optimize the code due to time constraints. We will make improvements in the future.

\section{RELATED WORK}

With the development of distributed systems, Nakamoto\cite{8} proposed to use blockchain to achieve decentralized currency. However,
the bitcoin script language is not a Turing-complete program language, it can only be used as a decentralized currency. But the
design of the decentralized management system can give us a lot of
revelations about one possible application of distributed networks.
Then, Vitalik Buterin\cite{2} proposed a new decentralized network,
Ethereum. It is not only a decentralized currency but also offers a
platform for us to write some rules. A smart contract is a collection
of self-verifying, self-executing, and tamper-resistant algorithms\cite{7}.
Because the smart contract is some code run on the blockchain, it
will not be modified and all the rules are published. When we do
not want to trust others, we can use the smart contract to write the
code and run the code. The code will be managed by most nodes.
A Turing-complete programming language is required to create
an expressive customized smart contract. The Eth smart contract’s
code is written in a stack-based bytecode language and executed
in the Ethereum Virtual Machine (EVM). Ethereum platform contracts may be written in a variety of high-level languages, including Solidity and Serpent\cite{12}. The applications of the Ethereum can be expanded to the machine learning fields \cite{2230,2233,2235} and data analysis \cite{2231,2232,2234}.

We usually use the smart contract to define some rules that involve a lot of people who don’t know each other well. Using the
smart contract, M et al.\cite{11} suggested a novel scalable and distributed data sharing system that incorporates access control for
smart agriculture. This means that the combination of the Internet
of Things and smart contracts has a wide variety of potential applications in the area of smart agriculture. Wang et al.\cite{13} suggested a
secure and auditable private data sharing (SPDS) method in a smart
grid data processing-as-a-service mode. They built a new structure
using blockchain technology, and smart contracts are used to set
fine-grained data usage regulations. The technology is designed
to assess the security of private data in an intelligent power grid.
Patient privacy must also be safeguarded in the medical area. Medical images account for nearly 70\% of medical diagnostic data, but
the traditional policy of data protection which shares the managed
video resources within the organization cannot control the medical
information at the time of treatment. Therefore, Tang et al.\cite{10}
propose a secure sharing method of medical images based on the
blockchain smart contract and credit scores. They have achieved a
medical image sharing system that can be trusted and supervised
across organizations and regions through a distributed, reliable
database blockchain that records the image sharing process. In addition, Zhang et al.\cite{14} introduce evaluation metrics to analyze
the smart contract code in the health domain. Smart contracts are
also widely used in the field of renting or sharing items. Zhou
et al.\cite{15} presented a blockchain-based decentralized car-sharing
management mechanism that makes use of smart contracts. The
enormous base stations of the Internet of Vehicles dispersed across
a large region are utilized to collaborate on the development of a
distributed system using blockchain to replace insecure third-party
servers. In a decentralized technology, the access control procedure
may be conducted automatically by any base station using smart
contracts. The system provides a safe framework for exchanges
between cars, people, and application providers, hence eliminating
security concerns. Sara\cite{9} highlighted the coupling of supply chain
and smart contracts as a potential application opportunity for smart
contracts. The key argument for using blockchain in supply chain
applications is the immutability of blockchain data. The benefits
of employing distributed ledger technology in the context of the supply chain include tracking and monitoring the processes from
product production to distribution, assuring quality control, and
offering an integrable and trustworthy process.
\section{CONCLUSION}
We build our application on Ethereum using the smart contracts
technology.Our application allows agents to upload used car information and buyers to auction off vehicles to transfer ownership.
It brings great convenience to buyers and owners, and the use of
smart contracts improves the security and privacy of the used car
bidding process.

\end{document}